\documentclass[%
showpacs,preprintnumbers,
 amsmath,amssymb,
 aps,
 pre,
 lengthcheck,%
]{revtex4-1}

\usepackage{amsmath,amssymb}
\usepackage{graphicx}
\usepackage{dcolumn}
\usepackage[latin1]{inputenc}
\usepackage{bm}
\usepackage{color}

\begin{document}

\preprint{APS/123-QED}

\title{Measuring degree-degree association in networks}

\author{Mathias Raschke, Markus Schläpfer and Roberto Nibali}
\affiliation{\mbox{Laboratory for Safety Analysis, ETH Zurich, CH-8092 Zurich}}

\date{\today}

\begin{abstract}
The Pearson correlation coefficient is commonly used for quantifying the global level of degree-degree association in complex networks. Here, we use a probabilistic representation of the underlying network structure for assessing the applicability of different association measures to heavy-tailed degree distributions. Theoretical arguments together with our numerical study indicate that Pearson's coefficient often depends on the size of networks with equal association structure, impeding a systematic comparison of real-world networks. In contrast, Kendall-Gibbons' $\tau_{b}$ is a considerably more robust measure of the degree-degree association.
\end{abstract}


\maketitle
\vspace{-0.2cm}
\section{Introduction}
\label{sec:intro}
\vspace{-0.2cm}
In many scientific fields, ranging from biology to sociology and engineering, network studies have recently brought substantial insights into the underlying connectivity patterns of different systems \cite{doro2003evolution,Barratt:2008,Castellano:2009}. Going beyond characterizing the network topology by the essential degree distribution \cite{citeulike:90557}, extensive research has focused on degree-degree correlations \cite{Newman:2002,Newman03thestructure,Barratt:2008}. A positive degree-degree correlation implies that nodes with a similarly small or large degree tend to be connected to each other. A negative degree-degree correlation accordingly indicates that the nodes tend to be connected to nodes with a considerably different degree. In the statistical physics community, the global level of correlation is commonly quantified by the Pearson coefficient $r$ \cite{Newman:2002}. However, there is a substantial drawback of this measure; its value strongly depends on the network size and might even vanish for large networks, as recently shown in \cite{Dorogovtsev:2010, Menche:2010}.

In this Brief Report, we apply bivariate distributions and different association measures from the field of statistics to describe and quantify the structure of networks. 
The term ``association'' (or ``dependency'') as used here refers to the general relation between two random variables \cite{Upton:2008}, while the term ``correlation'' is restricted to a single measure \cite{Mari:2001}. Both theoretical arguments and our numerical study indicate that, regarding the size of networks with a heavy-tailed degree distribution, Kendall-Gibbons' $\tau_{b}$ is a considerably more robust association measure than the Pearson coefficient $r$. 

In the following, we introduce the concept of the probability matrix as an application of bivariate distributions to networks. We then provide an overview of important association measures for bivariate distributions. Their suitability to quantify degree-degree associations in heavy-tailed networks is discussed in theory and numerically investigated.
\vspace{-0.2cm}
\section{Representing degree-degree association by a probability matrix}
\label{sec:probmatrix}
\begin{figure}[b!]
  \centering
  \includegraphics*[scale=0.42]{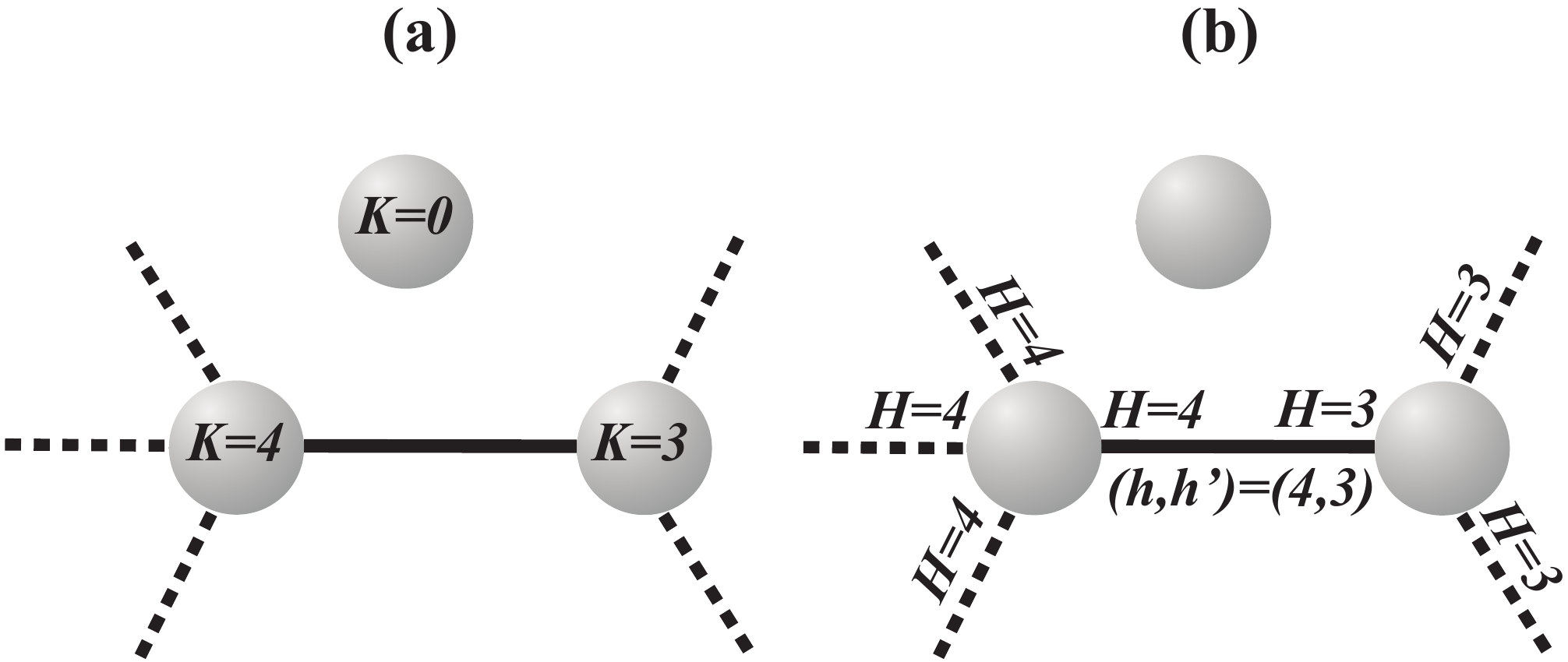}
  \caption{Different assignment of the number of edges. (a) $K$ edges per node, (b) $H$ edges connected to the end of an edge.}
  \label{fig:fig1}
\end{figure}

\vspace{-0.2cm}
The total number of edges being connected to a node is usually called degree and is denoted by $k$. However, the adequate representation of the degree-degree association requires the distribution of the number of edges $h$ connected to the end of an edge, including the considered edge itself. These integer numbers are realizations of the random variables symbolized with the capitals $K$ and $H$. It is $H=K$ for a specific node, except for $K=0$. The different assignment is illustrated in Fig.$~\ref{fig:fig1}$.
We denote by $P_{k}(k)$ the distribution function of $K$, which alternatively can be written as $P_{k}(K=k)$ and corresponds to the probability that a random variable $K$ has the value of realization $k$. The random variable $H$ has the distribution function $P_{h}(h)$. The functions $P_{k}(k)$ and $P_h(h)$ are related by
\begin{equation}
\label{eq:1}
P_{h}(h)=P_{k}(h)h/\langle k \rangle,
\end{equation}
\begin{equation}
\label{eq:2}
\langle k \rangle < \langle h \rangle,
\end{equation}
where $\langle k \rangle$ is the expectation of $K$, being different to the expectation $\langle h \rangle$ of the random variable $H$. 

Each edge has a realized pair of $H$, as depicted in Fig.~\ref{fig:fig1} with the exemplary pair $(h,h')=(4,3)$. The distribution of these pairs can then be described by the ``probability matrix'' $P(h,h')$, which corresponds to the joint distribution introduced in \cite{weber:046111}. The probability matrix differs to common bivariate discrete distributions of random variables, as each realization of $H$ occurs $H$ times at a node and is in $H$ pairs $(H,H')$. Furthermore, each pair $(H,H')$ has a twin $(H',H)$ in case of a network with undirected edges. Nevertheless, we treat $P(h,h')$ as a common bivariate discrete distribution, assuming that the differences are negligible. It should be noted that the balance condition formulated in \cite{boguna-2002-66} still holds, whereby the distinction between $h$ and $k$ allows describing the relations given by Eqs. (\ref{eq:1}) and (\ref{eq:2}) in a statistically correct manner.
\section{Association measures for discrete random variables}
\label{sec:association}
Different measures of association are presented here for general pairs of discrete random variables $(X,Y)$, as they are related to any bivariate distribution. These variables correspond to $(H,H')$ in networks (see Fig.~\ref{fig:fig1}). The association of a given pair $(X,Y)$ is fully described by its bivariate distribution $P(x,y)$, which is not directly comparable to the association of other pairs of random variables. Therefore, different measures have been developed allowing to compare two different association structures by a simple inequation. Each measure is based on a distinct definition, fulfilling a set of properties \cite{Mari:2001}. Firstly, the association measure is normalized so that its absolute value is not larger than 1. Secondly, the measure should be 0 if the two variables are independent and the absolute value should be equal to 1 in order to denote perfect association. Lastly, every observation in a sample should have the same weight in the computation of an association measure.

The commonly used measure for the degree-degree association in networks is the Pearson coefficient $r$ \cite{Newman:2002,Mari:2001}, written in its general form as
\begin{equation}
\label{eq:pearson}
  \rho_{p} = \frac{\langle XY\rangle -\langle X\rangle \langle Y\rangle}
                  {\sqrt{\left(\langle X^{2}\rangle -\langle X\rangle ^{2} \right)\left(\langle Y^{2}\rangle -\langle Y\rangle ^{2} \right)}}.
\end{equation}
The association is said to be positive if $r>0$ and negative if $r<0$. Pearson's coefficient is estimated by replacing the moments of $X$ and $Y$ with their estimations. The confidence interval for an estimation of $r$ depends on the distribution type of $X$ and $Y$.

Another important association measure is Spearman's rank correlation coefficient $\rho_{s}$. For continuous random variables $\rho_{s}$ is formulated similarly to $\rho_{p}$ by replacing in Eq. (\ref{eq:pearson}) the variables $X$ and $Y$ with $F(X)$ and $F(Y)$, being the cumulative distribution functions (CDF). In order to estimate $\rho_{s}$ for discrete random variables the CDF are replaced by the rank numbers of the ordered samples of $X$ and $Y$, respectively. An ordered sample of $n$ observations $X_i$ is $(X_1\leq X_2\leq...\leq X_i\leq...\leq X_n)$.

Kendall-Gibbons' $\tau_{b}$  \cite{kendall90rank} and Goodman-Kruskal's $\gamma_g$ \cite{goodman:1954} are two important association measures based on concordance. Two pairs $(X,Y)$ and $(X',Y')$ are concordant if $X > X'$ and $Y > Y'$, or $X < X'$ and $Y < Y'$. The pairs are dis-concordant if $X > X'$ and $Y < Y'$, or $X < X'$ and $Y > Y'$. Pairs with equal $X$ and pairs with equal $Y$ are called tied pairs. Kendall-Gibbons' $\tau_{b}$ for a sample of discrete random variables $(X,Y)$ is given by
\begin{equation}
\label{eq:kendall}
\tau_{b} = 2(n_c - n_d)/\sqrt{\left[n(n-1)-\xi_{x}\right][n(n-1)-\xi_{y}]},
\end{equation}
where $\xi_{x}\,=\,\sum_{i} m_x (i) (m_x (i)-1)$ with  $m_x (i)$ as the number of observations $X=i$. Accordingly, $\xi_{y}\,=\,\sum_{i} m_y (i) (m_y (i)-1)$ with $m_y (i)$ as the number of observations $Y=i$. The term $n_c$ denotes the number of concordant pairs
\begin{equation}
\label{eq:nc}
  n_{c} = \sum_{x=x_{\min}}^{x_{\max}-1}\sum_{y=y_{\min}}^{y_{\max}-1}
  				\left\{m_{x,y}\sum_{x'=x+1}^{x_{\max}}\sum_{y'=y+1}^{y_{\max}}m_{x',y'}\right\},
\end{equation}
and $n_d$ stands for the number of dis-concordant pairs
\begin{equation}
\label{eq:nd}
  n_{d} = \sum_{x=x_{\min}+1}^{x_{\max}}\sum_{y=y_{\min}+1}^{y_{\max}}
  				\left\{m_{x,y}\sum_{x'=x_{\min}}^{x-1}\sum_{y'=y_{\min}}^{y-1}m_{x',y'}\right\},
\end{equation}
where $m_{x,y}$ denotes the number of observations with $X=x$ and $Y=y$. The symbols $x'$ and $y'$ are written for $x$ and $y$ in order to distinguish the different pairs of observations. 
If the bivariate distribution $P(x,y)$ with its marginal distributions $P_x(x)$ and $P_y(y)$ is given, $\tau_{b}$ can be calculated as
\begin{equation}
\label{eq:kendall2}
\tau_{b} = \frac{\displaystyle{2(n_{c}-n_{d})}}%
  					{\displaystyle{\sqrt{\left(1-\sum_{x} P_{x}(x)^{2}\right)\left(1-\sum_{y} P_{y}(y)^{2}\right)}}},        
\end{equation}
where $n_c$ and $n_d$ are derived by Eqs. (\ref{eq:nc}) and (\ref{eq:nd}) after replacing $m_{x,y}$ by $P(x,y)$. The bivariate distribution, in turn, can be estimated as
$\hat{P}(x,y)=m_{x,y}/n$. The estimation of Kendall-Gibbons' $\tau_b$ by using Eq. (\ref{eq:kendall}) is equivalent to using Eq. (\ref{eq:kendall2}). Modifications of $\tau_b$ (e.g., \cite{Stuart1953,Somers1962}) are not considered in the scope of this Brief Report. Using the same notation, Goodman-Kruskal's $\gamma_g$ is calculated as
\begin{equation}
\label{eq:goodman}
\gamma_g =(n_{c}-n_{d})/(n_{c} +n_{d}).
\end{equation}
It should be noted that in contrast to Pearson's coefficient the concordance based measures are independent of the marginal distributions $P_x(x)$ and $P_y(y)$ \cite{BeirlantWolfe:2004}. Hence, a transformation of the random variables does not influence the value of $\tau_b$ and $\gamma _g$. Further association measures are given in \cite{Mari:2001}.
\section{Applicability to networks}
\label{sec:applicability}

The applicability to networks is discussed only for Kendall-Gibbons' $\tau _b$ and for the Pearson coefficient $r$. Spearman's $\rho_{s}$ is not further considered here as the rank of $H$ with $H$=1,2,3,... is equal to $H$ or is a linear transformation thereof. Hence, the values of $\rho_{s}$ do not significantly differ from those of $r$. Goodman-Kruskal's $\gamma _g$ does not consider 
all observations $(h,h')$ with the same weight (tied pairs are neglected) and therefore is not an adequate measure for the degree-degree association. 

The variable $H$ of real-world networks often follows a heavy-tailed distribution \cite{Barratt:2008}. The typical discrete distribution exhibiting such heavy-tail characteristics is the Zipf distribution. Its probability function is written as \cite{Gentle2005}
\begin{equation}
P(h) = h^{-\gamma}/\sum\limits_{h=1}^{h_{max}}h^{-\gamma}, h>0, \gamma>0,
\end{equation}
where the maximum degree $h_{max}$ can be both bounded or unbounded. The Zipf distribution is the discrete case of the continuous Pareto distribution \cite{Gentle2005}. Both distributions have no variance and second moment $\langle h^2 \rangle$ for $\gamma \leq 2$. Additionally, if the exponent is $\gamma \leq 1$ there is no expectation $\langle h\rangle$. Therefore, the theoretical arguments against the application of the Pearson coefficient $r$ to continuous heavy-tailed distributions (e.g., \cite{embrechts, RePEc:wbs:wpaper:wp01-13}) hold likewise for discrete heavy-tailed distributions. For instance, $r$ is not defined for $\gamma \leq 2$. In contrast, Kendall-Gibbons' $\tau _b$ is independent of the marginal distributions (see Sec. \ref{sec:association}) and hence independent of $\gamma$. Furthermore, Lindskog \cite{Lindskog:2000} stated that for the class of elliptical distributions the estimation of $r$ has a lower performance than the estimation of Kendall's $\tau$ (being the continuous case of Kendall-Gibbons' $\tau _b$), which indicates another drawback of Pearson's coefficient. Furthermore, the confidence interval for an estimation of $r$ again depends on the marginal distribution, whereas the confidence intervals for an estimation of $\tau _b$ can be constructed universally \cite{Samra:1988}.

We complement the theoretical discussion by numerically studying the behavior of both the Pearson coefficient $r$ and Kendall-Gibbons' $\tau _b$ with respect to heavy-tailed distributions of $H$. Thereby, a focus is set on the dependence of the measures on the maximum degree $h_{max}$, while maintaining the association structure. With an increasing $h _{max}$ the measures should converge to a constant value. The convergence behavior thus can be taken for a practical evaluation of the fitness of the association measures. Three symmetric bivariate distributions are used for generating the probability matrix $P(h,h')$. The first type is the mixed truncated Zipf distribution, which represents a mixture of the independent and the perfectly associated case of the Zipf distribution. The independent case is \cite{Mari:2001}
\begin{equation}
P_{1}(h,h')=P_{h}(h)P_{h}(h')=(hh')^{-\gamma}/\sum\limits_{h'=1}^{h_{max}}\sum\limits_{h=1}^{h_{max}}(hh')^{-\gamma},
\end{equation}
and the perfectly associated case is
\begin{equation}
P_{2}(h,h') = \left\{%
	\begin{array}{ll} 
		h^{-\gamma}/\sum\limits_{h=1}^{h_{max}}h^{-\gamma} 	& \text{for~}h=h' \,,\\[4ex]
		0 																						  		& \text{for~}h\ne h' \,.
	\end{array}\right.
\end{equation}
The mixing of these two cases is then given by
\begin{equation}
\label{mixedZipf}
P(h,h') = \frac{P_{1}(h,h')t+P_{2}(h,h')(1-t)} {\sum\limits_{h=1}^{h_{\max}}\sum\limits_{h'=1}^{h_{\max}}P_{1}(h,h')t+P_{2}(h,h')(1-t)},   
\end{equation} 
with the mixing function $t  = a\max(h,h')^{-c}$ and  $0 < a < 1, c \geq 0$. If $c=0$, then the mixing is simple and the Pearson coefficient becomes $r=1-a$ \footnote{Note that the mixing of the perfectly associated distribution with the independent distribution does not necessarily imply the mixing of a network with a perfect association (which does not exist) and a network with an independent degree-degree structure.}. For the second type the multivariate Zeta distribution according to Yeh \cite{Yeh2002} is used, with its survival function $\overline{F}$
\begin{equation} 
\overline{F}(h,h')=(1+(\alpha(h-1))^{\psi} +(\alpha(h'-1))^{\psi} )^{\eta}, h\geq1.
\end{equation} 
The probability matrix can thus be written as
\begin{equation} \label{Zeta}
P(h,h')=1-\overline{F}(h,1)-\overline{F}(1,h')+\overline{F}(h,h'),        
\end{equation} 
being normalized for the truncated case by the sum of $P(h,h')$ with $h \leq h_{max}$. For the last type we apply a continuous Pareto distribution for the marginal distributions and a negatively associated bivariate exponential distribution \cite{Gumbel:1960} to construct (negative) association structures. The resulting cumulative distribution function $F$ is formulated as
\begin{eqnarray} \label{pareto}
F(h,h') & = & 1-(h+1)^{-\gamma}-(h'+1)^{-\gamma} {} \nonumber\\ 
&& + ((h+1)(h'+1))^{-\gamma}e^{-\gamma^2 \theta ln(h+1)ln(h'+1)}{} \nonumber\\   
\end{eqnarray}
and $P(h,h')$ can then be calculated as
\begin{eqnarray} \label{pareto_P}
P(h,h') & = & F(h,h')+F(h-1,h'-1)  {} \nonumber\\
&& -F(h,h'-1)-F(h-1,h'),    
\end{eqnarray}
being normalized again as $h \leq h_{max}$, so that the sum of $P(h,h')$ equals 1. 

\begin{figure}[t!]
\vspace{-0.2cm}
\includegraphics*[scale=0.54]{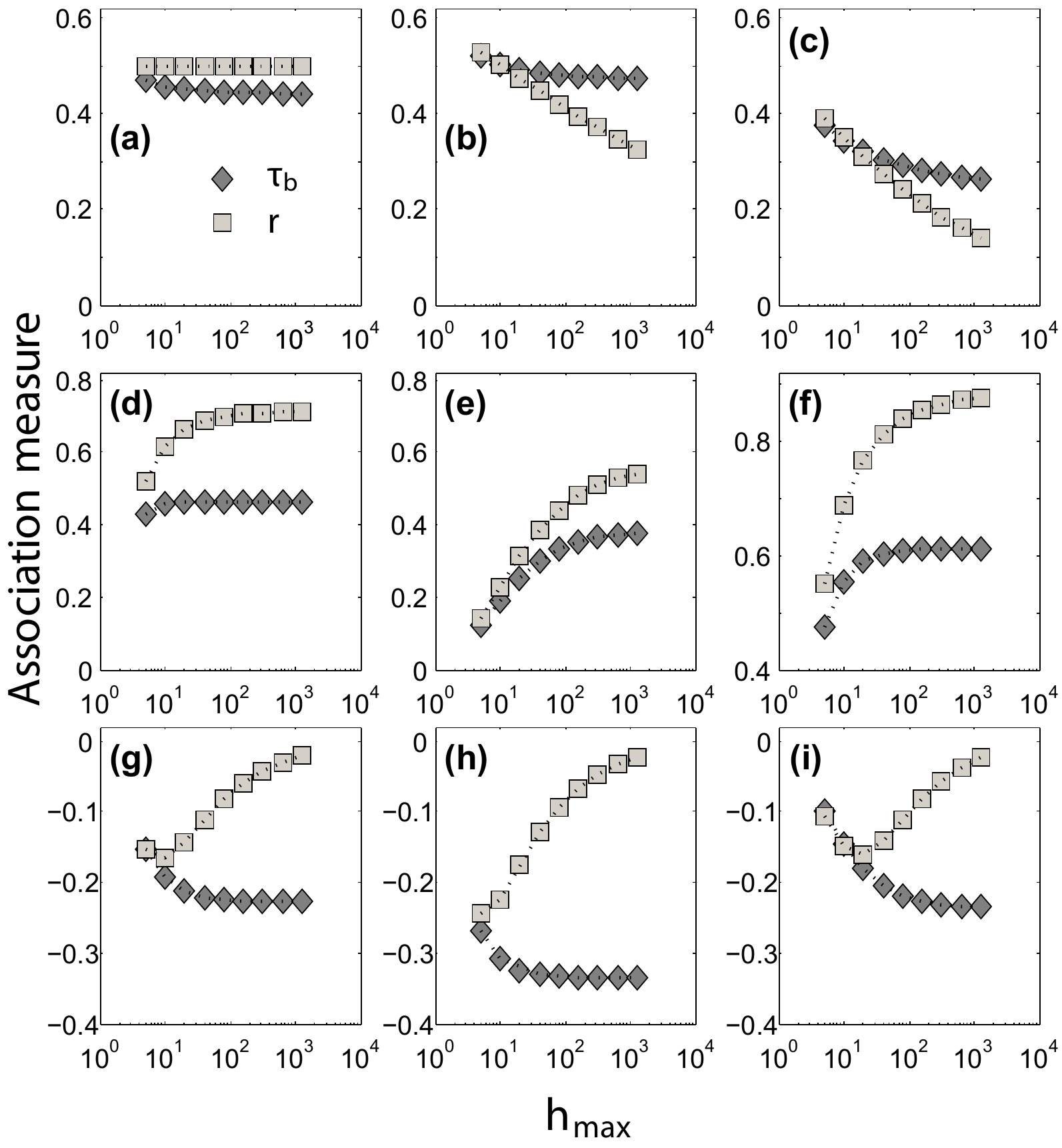}
\vspace{-0.65cm}
  \caption{\label{fig:fig2}Values of the association measures $r$ (squares) and $\tau_{b}$ (diamonds) in relation to $h_{max}$. (a) Zipf distribution with $a=0.5$, $ c=0$, $ \gamma=1.5$, (b) Zipf distribution with $a=0.6$, $ c=0.1$, $ \gamma=2$, (c) Zipf distribution with $a=0.5$, $ c=0.2$, $ \gamma=1.5$, (d) Zeta distribution with $\alpha=10$, $\psi=3$, $\eta=1$, (e) Zeta distribution with $\alpha=10$, $ \psi=1$, $\eta=1$,
(f) Zeta distribution with $\alpha=10$, $ \psi=3$, $\eta=0.5$, (g) Pareto distribution with $ \theta=0.5$, $\gamma=1.5$, (h) Pareto distribution with $ \theta=1$, $ \gamma=1.5$ and (i) Pareto distribution with $ \theta=0.5$, $ \gamma=1$. The dotted lines serve as a guide to the eye.}
\vspace{-0.65cm}
\end{figure}

Based on these three types of probability matrices, the Pearson coefficient $r$ and Kendall-Gibbons' $\tau _b$ have been determined for different distribution parameters and different maximum degrees $h_{max}$. The resulting values are reported in Fig. \ref{fig:fig2}. Kendall-Gibbons' $\tau _b$ converges faster to a constant value in case of the positive association of both the Zipf and the Zeta distribution [Fig. 2 (b)-(f)]. Pearson's coefficient is only slightly more stable in the specific case of the Zipf distribution with $c=0$ [Fig. 2 (a)]. In case of the negative association given by the probability matrix based on the Pareto distribution [Fig. 2 (g)-(i)] Kendall Gibbons $\tau_b$ is again significantly more stable than the Pearson coefficient $r$, which shows a non-monotonic behavior [e.g., Fig. 2 (i)] and converges to zero with increasing $h_{max}$. This adverse behavior of $r$ further confirms the results recently presented in \cite{Dorogovtsev:2010, Menche:2010}. Hence, our numerical study indicates that Kendall-Gibbons' $\tau_b$ is a more reliable association measure than the widely used Pearson coefficient $r$ due to its higher robustness with respect to changes in network size.
\section{Conclusions}
\label{sec:conclusions}
In this Brief Report, different measures for degree-degree association have been discussed with respect to their applicability to networks with a heavy-tailed degree distribution. Thereby, the network structure has been represented by a probability matrix allowing to distinguish between the degree related to a given node and the degree related to the end of a given edge. Theoretical arguments together with our numerical study confirm that the widely used Pearson coefficient is hardly feasible to quantify and compare degree-degree associations in heavy-tailed networks, mainly due to its strong size-dependence. Our findings indicate that Kendall-Gibbons' $\tau_{b}$ is a significantly more robust measure with respect to the size of networks with equal association structure.
\begin{acknowledgments}
M.R. acknowledges ``swisselectric research'' and the Swiss Federal Office of Energy (project No. V155269) for co-funding the present work.
\end{acknowledgments}
%
\bibliography{Master}
\end{document}